\documentclass{elsart}
\usepackage{amssymb}

\usepackage{epsfig}

\begin{document}

\begin{frontmatter}

\today

{\bf \Large Multi-Terminal Superconducting Phase Qubit}

\author{\small M.H.S. Amin$^a$, A.N. Omelyanchouk$^{b}$, A. Blais$^c$,
Alec Maassen van den Brink$^a$,}
\author{\small G. Rose$^a$, T. Duty$^a$, and A.M. Zagoskin$^{a,d}$}

\address{$^a$D-Wave Systems Inc., 320-1985 West Broadway,
Vancouver, BC, V6J 4Y3, Canada; $^b$B.I.Verkin Institute for Low
Temperature Physics and Engineering, Ukrainian National Academy
of Sciences, Lenin Ave.\ 47, Kharkov 310164, Ukraine; $^c$Centre
de Recherche sur les Propri\'et\'es \'Electroniques de
Mat\'eriaux Avanc\'es and D\'epartement de Physique, Universit\'e
de Sherbrooke, Sherbrooke, Qu\'ebec, J1K 2R1, Canada; $^d$Physics
and Astronomy Dept., The University of British Columbia, 6224
Agricultural Rd., Vancouver, BC, V6T 1Z1, Canada}

\begin{abstract}
Mesoscopic multi-terminal Josephson junctions are novel devices
that provide weak coupling between several bulk superconductors
through a common normal layer. Because of the nonlocal coupling
of the superconducting banks, a current flow between two of the
terminals can induce a phase difference and/or current flow in the
other terminals. This ``phase dragging" effect is used in
designing a new type of superconducting phase qubit, the basic
element of a quantum computer. Time-reversal symmetry breaking
can be achieved by inserting a $\pi$-phase shifter into the flux
loop. Logical operations are done by applying currents. This
removes the necessity for local external magnetic fields to
achieve bistability or controllable operations.
\newline
\newline
{\em Keywords:} Quantum Computing, Qubit, Multiterminal, Josephson
junction

\end{abstract}

\end{frontmatter}

\small

Although time-domain coherent oscillations have been observed in
superconducting charge qubits \cite{nakamura}, the short decoherence time $%
\tau _{\varphi }$, due to the fluctuations of the background charges,
prevents these qubits from being a good candidate for large-scale quantum
computing. Phase qubits, on the other hand, can couple weakly to the
background charges and therefore potentially have larger $\tau _{\varphi }$.
To achieve a reasonably long $\tau _{\varphi }$, it is necessary to have a
``quiet''~\cite{ioffe} phase qubit---with small magnetic coupling to the
environment, or equivalently, small inductance. A usual rf-SQUID can show
bistability only when the inductance $L$ of the ring exceeds $2\pi \Phi
_{0}/I_{c}$ \cite{mss}, and therefore cannot be quiet. Here, $I_{c}$ is the
Josephson critical current of the junction and $\Phi _{0}=h/2e$ the flux
quantum. To overcome this problem, three Josephson junctions have been
included in a superconducting ring \cite{mooij}. One of the three Josephson
phases is fixed by the other two and the external flux, which leaves the
SQUID with two degrees of freedom, making bistability possible even when $L=0
$.

\begin{figure}[ht]
\epsfysize 7cm \epsfbox[-100 180 400 600]{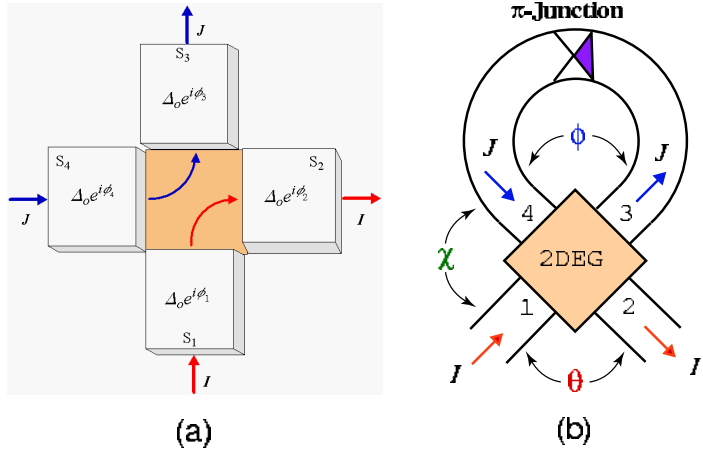}
\caption[]{\small Mesoscopic four-terminal junction (a) and
four-terminal SQUID (b). The $\pi$-junction is included in the
ring to attain bistability.} \label{fig1}
\end{figure}

A four-terminal junction is described by three phase variables (see below).
Connecting two of the terminals by a superconducting ring will fix one of
the phases to the external flux (neglecting the inductance of the loop, $L$%
). The resulting four-terminal SQUID \cite{oo} will have two degrees of
freedom and can exhibit bistability at small $L$. Bistability of a
four-terminal SQUID made from microbridges has been observed experimentally
\cite{vleeming}. As we shall see, with a mesoscopic 4-terminal junction
(Fig.\ 1a) it is possible to have bistability even at $L=0,$ due to the
phase-dragging effect \cite{zo,AOZ}, .

A mesoscopic 4-terminal junction is shown in Fig.\ 1a. The four bulk
superconductors are connected to each other via a 2-dimensional electron gas
(2DEG) region. The phase of the order parameter in the $i$-th terminal is
denoted by $\phi_i$. When the dimensions of the 2DEG region are smaller than
the superconducting coherence length in the banks, the total current $I_i$
flowing into the $i$-th terminal depends on the superconducting phases $%
\phi_j$ in all the banks through~\cite{zo}

\begin{equation}
I_{i}= \frac{\pi \Delta_0 }{e} \sum_{j=1} ^{4} \gamma_{ij} \sin \frac{%
\phi_{ij}}{2} \tanh{\left[ \frac{\Delta_0}{2T} \cos \frac{\phi_{ij}}{2} %
\right] },  \label{Ii}
\end{equation}
where $\gamma_{ij}$ are Josephson coupling constants \cite{note},
$\phi_{ij}\equiv \phi_{i}-\phi_{j}$, and $\Delta_0$ is the superconducting
gap. We study the system at temperatures close to $T=0$,\linebreak where
decoherence due to the environment is minimal. In this limit, the Josephson
energy associated with the four-terminal junction is given by

\begin{equation}
\mathcal{E}_J \equiv {\frac{E_J }{E_0}} = - {\frac{1 }{\gamma_{12}}}
\sum_{i<j} \gamma_{ij} \left|\ \cos {\frac{ \phi_{ij} }{2 }} \right|.
\label{EJ0}
\end{equation}
Here, $E_0=\hbar I_0/e$ and $I_0= \pi {\gamma}_{12} \Delta_0/e$ are the
Josephson energy and critical current for the subjunction 1--2 at $T=0$,
respectively.

\begin{figure}[ht]
\epsfysize 5cm \epsfbox[-60 250 400 530]{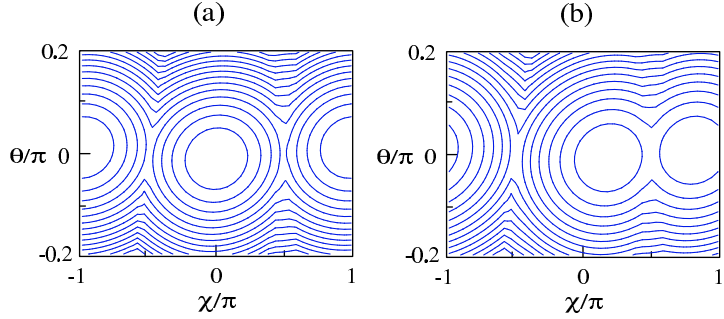} \caption[]
{\small Contour plot of the free energy at $T=0$ and $\mathcal{\
I}=0 $. (a) \ Junction with $\epsilon=0$. (b) \ Junction with
$\epsilon=0.03$. Other parameters are $\gamma=0.1,\
\delta_1=\delta_2=0.05$.} \label{fig2}
\end{figure}

A mesoscopic 4-terminal SQUID is constructed from the 4-terminal junction by
connecting two of the terminals via a superconducting ring (Fig.~1b, ignore
the $\pi$-junction for the moment). We label the terminals in such a way
that subjunction 1--2 forms the bias circuit carrying current $I$, and
subjunction 3--4 makes the flux loop with current $J$ and flux $\Phi$
threading the ring. We introduce new variables by $\phi_{1,2} = (\mp \theta
+ \chi)/2$ and $\phi_{3,4} = (\pm \phi-\chi)/2$, implicitly setting $\sum
\phi_i=0$, which is allowed because the overall phase is arbitrary. The
phase differences $\theta$ and $\phi$ are between terminals 1--2 and 3--4,
respectively. On the other hand, $\chi$ is the overall phase difference
between the ring and the bias circuit. It is also useful to define the new
dimensionless parameters

\begin{eqnarray}
\gamma&=& (\gamma_{13}+\gamma_{23}+\gamma_{14}+\gamma_{24})/\gamma_{12}
\nonumber \\
\epsilon&=&(\gamma_{13}+\gamma_{23}-\gamma_{14}-\gamma_{24})/\gamma_{12}
\nonumber \\
\delta_1&=&(\gamma_{13}-\gamma_{23}-\gamma_{14}+\gamma_{24})/\gamma_{12}
\nonumber \\
\delta_2&=&(\gamma_{13}-\gamma_{23}+\gamma_{14}-\gamma_{24})/\gamma_{12}
\nonumber \\
\kappa &=& \gamma_{34}/\gamma_{12}.  \label{params}
\end{eqnarray}

In general our system has a 3D phase space ($\phi,\theta,\chi$). However, we
are interested in the regime where $\kappa \ll \gamma \ll 1$ and $L
\rightarrow 0$, so that the self-generated flux by the ring be very small ($%
\propto \gamma I_0 L \ll \Phi_0$). Therefore, $\phi$ is practically fixed by
the external field and/or by a $\pi$-phase shifter inserted into the ring
(see below), and we can study the system in the 2D phase space of ($%
\theta,\chi$).

\begin{figure}[ht]
\epsfysize 5cm \epsfbox[-60 270 400 550]{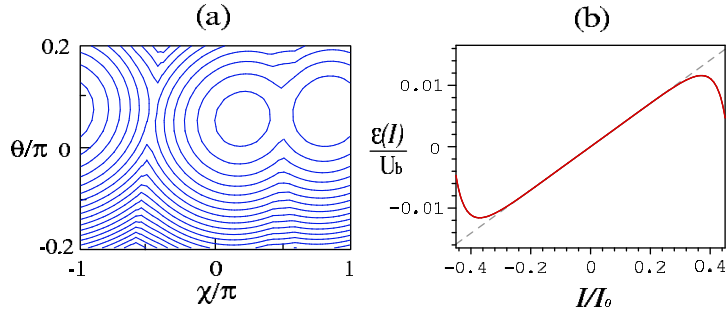}
\caption[]{\small  (a) Contour plot of the free energy for the
system of Fig.~2b at $\mathcal{I}=0.05$.\ (b) Solid line is the
energy bias as a function of the transport current, normalized to
the barrier height, for the system of Fig.~2b. Dashed line is the
linear approximation.} \label{fig3}
\end{figure}

Applying an external flux $\Phi _{e}=\Phi _{0}/2$ to the superconducting
ring makes the system bistable (as in the rf-SQUID or 3-junction cases),
meaning that the free energy of the system has two local minima,
corresponding to opposite directions of current in the ring. Note that the
external flux is not used to manipulate the flux (qubit) state. It therefore
can be fixed to $\Phi _{0}/2$ for all qubits. This opens the possibility of
replacingthe external fluxes by a $\pi $-phase shifter \cite{pijnc} in each
qubit's superconducting ring, as shown in Fig.~1b.  The net effect is the
same but this has the advantage that the $\pi $-phase shifter does not bring
in extra coupling to the electromagnetic environment. In the regimeof
interest ($L\rightarrow 0),$ $\phi =\pi $ and the free energy of the system
is given by

\begin{equation}
\mathcal{U}=-\mathcal{I}\theta +\mathcal{E}_{\mathrm{J}}(\theta ,\chi ,\phi
=\pi ),  \label{U}
\end{equation}
where $\mathcal{I}\equiv I/I_{0}$. Contour plots of this free
energy at two different sets of parameters are given in Figs.~2
and 3a. Tunneling between the potential wells is enabled by
charging effects, with the electrostatic capacitance of the
system defining the effective ``mass'' in the kinetic energy term
(see e.g. \cite{SCHOEN}, sec.2.2.2).

When $\mathcal{I}=0$, the two minima of $\mathcal{U}$ have equal energy.
Contour plots of the free energy of the system at $\mathcal{I}=0$ are shown
in Fig.~2. As is clear from the figure, with the parameters chosen, the
minima are located very close to $\theta =0$. In extended phase space, there
are also other minima, near $\theta =2\pi n$ ($n$ an integer). Those minima
are separated from the ones shown in the figure by haigh and wide potential
barriers. Therefore, tunneling in those directions is negligible. The
situation is different for $\chi $. When $\epsilon =0$, as is the case for a
system with a square 2DEG region and four equivalent terminals, the minima
are equidistant at $\chi =0$, $\pm \pi $, with equal barriers between them
(Fig.~2a). Therefore the tunneling probabilities in the left and right
directions are the same. This is undesirable for qubit application because
it makes the system sensitive to random charges in the environment \cite
{orlando:99,blatter:01}. However, making $\epsilon \neq 0$ will move two of
the minima closer together, making the barrier heights unequal (Fig.~2b).
Pairs of minima are then isolated, and one can associate a given pair of
minima with the logical qubit states $\{|0\rangle ,|1\rangle \}$. This
regime can be achieved easily by choosing a rectangular 2DEG region instead
of a square one \cite{long}.

Applying a nonzero transport current $\mathcal{I}$ moves the minima from
being centered around $\theta=0$ to some $\theta=\theta_0(\mathcal{I})$.
More importantly, it removes their degeneracy. Fig.~3a displays the contour
plot for $\mathcal{U}$ using the parameters of Fig.~2b, but with $\mathcal{I}%
=0.05$. As a result of the applied current, the two minima are now clearly
unequal.

The energy difference between the two minima $\varepsilon (\mathcal{\ I})$
is plotted in Fig.~3b. As is evident from the figure, this energy bias is
linearly dependent on $\mathcal{I}$ for a relatively wide range of the
transport current: $-0.3 \lesssim \mathcal{I} \lesssim 0.3$. We can
therefore approximate it by $\varepsilon (\mathcal{I})= \varepsilon_0
\mathcal{I}$, where $\varepsilon_0$ is given by \cite{long}

\begin{equation}
\varepsilon_0 = {\frac{(\gamma \delta_1 - \epsilon \delta_2) [\gamma
\epsilon (\delta_1^2 + \delta_2^2) + \delta_1\delta_2 (\gamma^2 +
\epsilon^2)] }{4(\gamma^2 + \epsilon^2)}}.  \label{ep0}
\end{equation}

To study the quantum dynamics of this system, we need to know the
capacitances between the terminals of the 4-terminal junction. In general,
there exists a capacitance between any two terminals of the system and one
has to find the component of the effective mass tensor along the direction
of tunneling in the same way as in Ref. \cite{orlando:99}. However, as is
clear from Fig.\ 2b, the difference in $\theta$ (and also $\phi$) from one
minimum to another is very small compared to that in $\chi$. The tunneling
is therefore effectively in the $\chi$ direction. Using a simplified 1D
model we find the tunneling matrix element at $\mathcal{I}=0$ to be $\Delta
\sim \hbar \omega_0 e^{- \sqrt{U_b/E_c}}$, where $\omega_0=
[8(\gamma^2+\epsilon^2)]^{1/4} \sqrt{E_0 E_c}$ is the plasma frequency at
the minima, $E_c=e^2/2C_{\mathrm{eff}}$ is the charging energy, $C_{\mathrm{%
eff}}$ is the effective capacitance in the direction of tunneling, and

\begin{equation}
U_b = {\frac{(\gamma - |\epsilon|)^2 E_0 }{2\left[ \gamma + |\epsilon| +
\sqrt{2(\gamma^2 + \epsilon^2)}\, \right]}}
\end{equation}
is the barrier height between the two nearest minima \cite{long}.

As mentioned above, an $\mathcal{I}\ne 0$ lifts the degeneracy between the
lowest-energy states and therefore stops the coherent tunneling. This energy
difference induces a relative phase between the logical states. Therefore,
control over the transport current suffices to manipulate the effective
one-qubit Hamiltonian $H_{\mathrm{eff}}=\Delta(I)\sigma_x +
E_0\varepsilon(I)\sigma_z$. Entangling operations between two qubits are
possible through voltage-controlled couplings provided by additional 2DEGs
\cite{long}. Combining the two regimes of zero and nonzero $\mathcal{I}$ and
using 2-qubit coupling, it is possible to perform any quantum gate
operations \cite{bz:2000}.

Most of the arguments about decoherence discussed in Refs. \cite{mss,tian}
carry over to this system. There are however also two sources of decoherence
different from those discussed in the references. The first is decoherence
caused by fluctuations of the transport current $\mathcal{I}$. This can be
reduced by increasing the internal resistance of the current source and
working at low temperatures \cite{mss,long}. Moreover, the current carried
by the quasiparticles through the normal region can also cause decoherence.
As shown in \cite{yeyati}, the quasiparticle (shunt) resistance is $R_{%
\mathrm{qp}} \propto T \cosh^2 (E_A / 2 k_B T)$, where $\pm E_A$ are the
energies of the Andreev bound states inside the normal region. To achieve
(exponentially) large $R_{\mathrm{qp}}$ and therefore long $\tau_\varphi$,
it is necessary to work at temperatures far below $E_A$. In systems with a
large 2DEG region, the energy scale $E_A$ is inversely proportional to the
dimensions of the normal region and can be much smaller than the gap $%
\Delta_0$ (which determines the energy scale in tunnel junctions). For short
junctions on the other hand (which is the case here), $E_A \sim \Delta_0
\cos (\Delta \phi /2 )$ and can be large if the phase difference $\Delta \phi
$ is not close to $\pi/2$. A phase-dependent conductance in agreement with
the above picture has been observed \cite{rifkin}.

In the limit studied in this paper, the time scale of the dynamics is set by
the Josephson and charging energies, as well as by the coupling
coefficients~(\ref{params}). For a junction size of 100 nm, we estimate $%
I_0\sim 10^{-7}$A \cite{heida:98} and $C_{\mathrm{eff}}\sim 10^{-13}$F.
Taking $\gamma=0.1$, $\delta_1=\delta_2=0.05$ and $\varepsilon=0.04$, we
obtain $\Delta\sim 0.1$GHz while tunneling through the barrier separating
the pairs of minima is $10^{-3}$ smaller. The dynamics is thus effectively
restricted to one pair of minima in phase space. Moreover, from (\ref{ep0})
we obtain $E_0\varepsilon_0\sim 0.01$GHz. Using the latter result, we
estimate that up to $10^{5}$ operations can be performed within the
decoherence time due to fluctuations of the transport current \cite{long}.

We would like to thank A.-M. Tremblay for stimulating discussions. AMB
thanks the Chinese University of Hong Kong for its hospitality. AB was
supported in part by NSERC, D-Wave Systems Inc. and FCAR.



\begin{thebibliography}{99}
\bibitem{note}  Analytical expressions for $\gamma_{ij}$ in the case of a
clean rectangular 2DEG region are given in Ref. \cite{zo}.
\end{thebibliography}
\end{document}